\documentclass[conference]{IEEEtran}
\IEEEoverridecommandlockouts

\usepackage{cite}
\usepackage{amsmath,amssymb,amsfonts}
\usepackage{algorithmic}
\usepackage{graphicx}
\usepackage{textcomp}
\usepackage{xcolor}
\usepackage{caption}
\usepackage{subcaption}
\usepackage{comment}
\usepackage{svg}
\usepackage[font={footnotesize}]{caption}
\usepackage{url}

\def\BibTeX{{\rm B\kern-.05em{\sc i\kern-.025em b}\kern-.08em
    T\kern-.1667em\lower.7ex\hbox{E}\kern-.125emX}}
\begin{document}

\title{Memory-less and Backscatter-less Tunnel Diode Harmonic Signatures for RFID\\
\thanks{\textcolor{red}{\textcopyright 2025 IEEE. Personal use of this material is permitted. Permission from IEEE must be obtained for all other uses, in any current or future media, including reprinting/republishing this material for advertising or promotional purposes, creating new collective works, for resale or redistribution to servers or lists, or reuse of any copyrighted component of this work in other works. Final version's URL: {https://ieeexplore.ieee.org/document/11082285} 
\\{DOI: 10.1109/JRFID.2025.3589528} }}
}

\author{Christopher Saetia,
 ~\IEEEmembership{Graduate Student Member, IEEE}, Kaitlyn Graves  ~\IEEEmembership{Graduate Student Member, IEEE}, \\ Serhat Tadik  ~\IEEEmembership{Graduate Student Member, IEEE},
Gregory D. Durgin, ~\IEEEmembership{Fellow, IEEE}\\
    \IEEEauthorblockA{School of Electrical Engineering, Georgia Institute of Technology, Atlanta, United States \\
    (csaetia3, kgraves30, serhat.tadik, durgin)@gatech.edu}
}

\maketitle

\begin{abstract}
Within the field of radio-frequency identification (RFID) research, tunnel diodes have traditionally been researched for extending backscatter read-ranges for ultra-high-frequency (UHF) RFID tags as reflection amplifiers due to their negative resistance. This same negative resistance can also be used to help construct oscillators. This paper further explores the use of tunnel diodes to make oscillators for harmonic RFID applications and the natural harmonics that arise when biasing these diodes within their negative differential resistance (NDR) regions and with no interrogating signal from a transmitting source, such as an RFID reader, to injection-lock these diodes. These harmonics are characterized for five tunnel diode boards, made with the same components and with each board's fundamental frequencies measuring at above -15 dBm at a biasing voltage of 200 mV when measured over-the-cable. The best DC-to-RF conversion efficiency achieved in this work was 30$\%$. The occurrence of harmonics from the tunnel diodes creates unique harmonic signatures for each board and demonstrates possible harmonic RFID applications that can help RFID readers discover and even identify RFID tags with backscatter-less and memory-less IDs generated by such tunnel diodes.

\end{abstract}

\begin{IEEEkeywords}
Tunnel diodes, harmonic RFID, harmonic signature, negative resistance, tunnel diode oscillator
\end{IEEEkeywords}

\section{Introduction}
Tunnel diodes are heavily-doped semiconductor devices that exhibit \textit{negative resistance} at low biasing voltages, typically starting below 300 mV. They are capable of producing reflection coefficient magnitudes greater than unity at low turn-on voltages. Thus, they are used to make the tunneling reflection amplifiers that have commonly been studied for extending the read-ranges of passive radio-frequency identification (RFID) tags \cite{Amato_Range, Cheng_Phased_Based_Localization, Eid_Tunnel_Diode, Saetia}. Tunnel diodes can and \textit{should} be used to address issues, other than path loss, that affect the backscatter link quality between tag and reader. For example, Eid et al. \cite{Eid_Tunnel_Diode} investigated leveraging tunnel diodes' low turn-on voltages and negative resistance, respectively, for rectification and compensating resistance losses in an oscillator design. Saetia and Durgin \cite{Saetia_APSK} have exploited tunneling reflection amplifiers to create multi-bit symbols for amplitude-phase shift keying (APSK). Self-jamming, identifying-tags in cluttered environments, etc. are some other issues in RFID systems that tunnel diodes can be used to help address. Specifically, this paper aims to highlight the characterization of tunnel diodes' natural resonance harmonics and investigate using those harmonics for \textit{memory-less} identification of tags, offering an alternative method to generate tags' IDs and sense the presence of tags in an area, all \textit{backscatter-less}.

\subsection{The Harmonic RFID Concept}

\begin{figure}[thbp!]
    \centering
    \includegraphics[width=0.95\linewidth]{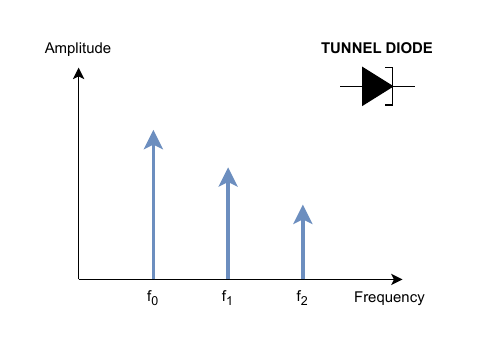}
    \caption{Tunnel diodes' nonlinearity allows them to produce resonating harmonics that can be used to help identify RFID tags that use them as backscatter modulators also.}
    \label{fig:TD_harmonics}
\end{figure}

Researchers have been exploring the use of frequency harmonics to better identify both chipless and chipped tag designs. Chipless RFID tags typically use different resonators to create a unique frequency response signature. As noted by Barbot et al. \cite{Barbot_Chipless_Read_Range}, these chipless tags are linear and time-invariant, making their read-ranges limited compared to traditional passive ultra-high frequency (UHF) RFID tags that are non-linear. Being \textit{non-linear} allows traditional UHF tags to generate harmonic and inter-modulated signals and, thus, backscatter information on frequencies outside the carrier frequency of the interrogating signal. This use of non-carrier frequencies improves these tags' detection since the reader's receiver can distinguish the tags' weaker backscattered signals from the directly interfering signal from the reader's transmitter and the other reflected signals from objects in the environment that match the carrier frequency. Barbot et al. \cite{Barbot_Chipless_Read_Range} noted that the inclusion of a nonlinear element to a chipless tag allows the tag to be nonlinear. The use of schottky diodes to allow chipless tags to backscatter at the second-order harmonic of the fundamental/interrogating signal has been demonstrated in \cite{Hemour_Chipless} and \cite{Palazzi_Paper_Substrate_Chipless}.

Traditional chipped passive UHF tags are non-linear due to:
\begin{enumerate}
    \item the non-linear circuit components of their energy harvesters and integrated circuits/chips
    \item the switching between different impedance states to perform load modulation for backscatter communications operations
\end{enumerate}

Both aspects have been investigated to improve identification of these tags. Barbot \cite{Barbot_Nonlinear_Modulation} explored having the RFID reader detect passive tags after interrogating their nonlinear chips and looking for the tags' backscattered frequency components that are generated around the interrogation carrier signal's frequency. Piumwardane et al. \cite{HarmonicID} varied duty cycles of the toggling of a tag's load modulator to create different harmonic/frequency signatures to assign tags with unique IDs. 

Tunnel diodes, like any non-linear device, produces harmonics. In 1963, Guttmann \cite{Harmonic_Tunnel_Diode_Generation} suggested the use of tunnel diodes as harmonic generators due to their nonlinearity and negative resistance/conductance.  More recently, Gumber et al. \cite{Harmonic_Hemour} demonstrated the ability for tunnel diodes to backscatter on a fundamental and second harmonic to help resolve the RFID reader's self-jamming problem through frequency multiplexing the transmit and receive operations. Inspired by that work, this research aims to further investigate the idea of creating a harmonic tunneling tag for identification. In contrast to \cite{Harmonic_Hemour}, emphasis was placed on measuring the tunnel diode's frequency response or natural resonance when only given a DC-bias; thus:
\begin{enumerate}
    \item No injection-locking or any intentional external RF interrogating signal was used on the designed tunnel diode boards. These boards simply served as oscillators with no backscattering involved.
    \item No matching networks were used to determine the frequencies these tunnel diodes would have high reflection gains and resonance strengths.
\end{enumerate}

Furthermore, this work aims to explore the possibility of these tunnel diodes' harmonics acting as a \textit{memory-less and backscatter-less signature} that allows for a RFID reader to detect the presence of tags and identify tags. Since tunnel diodes are well-established in research literature as highly-desired components to include for backscatter modulator designs, this paper further highlights their dual-purpose to create "harmonic signatures" for harmonic RFID applications being currently researched in the RFID field. This paper is structured to first introduce a circuit model for oscillation and harmonic generation. Then over-the-cable and wireless measured results from five tunnel diode boards will be presented. Harmonic RFID applications that will benefit from the use of tunnel diodes will be posed, along with future work at the conclusion of this paper.

\section{The Tunnel Diode Oscillator Model}
\begin{figure}[htbp!]
    \centering
    \includegraphics[width=0.95\linewidth]{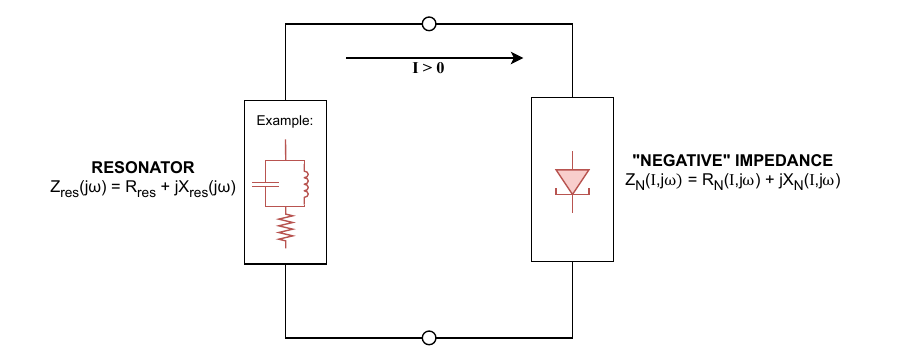}
    \caption{One-port negative resistance oscillator.}
    \label{fig:Neg_Impedance_Osc}
\end{figure}

In this work, the tunnel diodes are soldered onto printed circuit boards with a RF choke, a DC smoothing capacitor, and microstrip lines. Only a DC biasing voltage is needed to make these boards resonant, similar to how oscillators convert DC power to an AC output signal \cite{Pozar}. Thus, to the best of the authors' knowledge, the circuit model and reasoning for the frequency generation by this work's tunnel diode boards can be best explained by the well-established one-port oscillator model. Negative resistance is often used to create the one-port oscillator model, as depicted in Fig. \ref{fig:Neg_Impedance_Osc}. 

This negative resistance is important for creating sustained oscillations. A realistic passive resonant circuit by itself, such as an inductive-capacitive (LC) tank, will not create a continuous oscillation over time. The impedance losses will attenuate the strength of the resonator's output signal. An active, nonlinear device is needed to compensate for these losses in the circuit. Such device can provide \textit{negative resistance} for compensation.  Fig. \ref{fig:Neg_Impedance_Osc} shows a resonator load connected to a negative impedance device in a \textit{one-port} oscillator configuration. As described in \cite{Pozar}, the loop current, denoted by $I$, flowing through this oscillator circuit will be non-zero when the circuit behaves as an oscillator. The Kirchhoff voltage loop (KVL) expression for this circuit is then expressed in \eqref{eq:KVL_Osc} as the sum of the voltage drops across the resonator's impedance, $Z_{res}$, and negative impedance, $Z_{N}$.

\begin{equation}\label{eq:KVL_Osc}
    \left(Z_{res} (\omega) + Z_{N}(I,j\omega)\right)I = 0  \quad and \quad I>0
\end{equation}

The resonator's impedance $Z_{res}$ is a function of frequency ($\omega$), and the negative impedance $Z_{N}$ is a function of both loop current and frequency. Note, this dependence on current can be substituted with voltage, but for simplicity in presenting this oscillator model, current will be used.  As noted in \cite{Pozar}, from \eqref{eq:KVL_Osc}, a non-zero current forces the following conditions related to the real and imaginary components of the impedances to occur to satisfy the KVL loop: 

\begin{equation}\label{eq:cond1}
    R_{res} + R_{N}(I,j\omega) = 0
\end{equation}

\begin{equation}\label{eq:cond2}
    X_{res}(j\omega) + X_{N}(I,j\omega) = 0
\end{equation}

Since $R_{res}$ is a positive resistance value, then $R_{N}(I,j\omega)$ \textit{must} be less than zero, \textit{negative resistance}. The impedance values from the added $Z_N$ cancels out the impedance values of $Z_{res}$, allowing for a non-decaying current flow. Importantly, due to the impedances' dependence on frequency and current, specific frequencies and voltages/currents allow for conditions \eqref{eq:cond1} and \eqref{eq:cond2} to be met to ensure continuous oscillation. For steady-state and single-frequency stability where $\omega=\omega_{o}$, the frequency of steady oscillation, the Kurokawa’s condition, fully derived in \cite{Pozar}, helps ensure the oscillator maintains a stable resonance at the steady-state frequency, despite perturbation or changes to the circuit's frequency or current. Because tunnel diodes, like any diodes, are non-linear devices, harmonics of the fundamental resonating frequency should appear on the devices' measured output spectrum, unless filtered out.

Tunnel diodes, in their current-voltage curves, have negative differential resistance (NDR) regions that can provide the negative resistance in this one-port oscillator model at specific voltages. \cite{Kim_Tunnel_Oscillator, Eid_Tunnel_Diode, Wang2012, Varshney_Tunnel_Emitter, Varshney_TUnnelScatter} demonstrated the construction of tunnel diode oscillators. Specifically, Eid et al. \cite{Eid_Tunnel_Diode} and Varshney et al. \cite{Varshney_TUnnelScatter} both demonstrated the capability to use a tunnel diode as both a low-power reflection amplifier and oscillator for low-power wireless communications. Additionally, Varshney and Corneo \cite{Varshney_TUnnelScatter} used a simple tunnel diode oscillator that is injection-locked by a nearby backscattered signal from a tag to relay a stronger response to a receiver. For $\geq$100 GHz and even THz communications, research has been conducted to build high-frequency, resonant-tunneling-diode (RTD) oscillators that can output tens of micro-watts \cite{Yu_RTD} to greater than half a milli-watt\cite{spudat_RTD, Khalidi_RTD} of RF power. Even amateur hobbyists have used old Russian tunnel diodes to build simple tunnel diode oscillators. For example, \cite{Tunnel_Diode_Oscillator_Video} showcased a tunnel diode oscillator built by simply looping capacitors and a wire inductor around a tunnel diode's leads, making a harmonic-generating and tunable oscillator that is dependent on the biasing voltages. This same tunable and harmonic-generating behavior will be observed in this work.

\section{Board Design}

\begin{figure}[htbp!]
    \centering
    \includegraphics[width=0.85\linewidth]{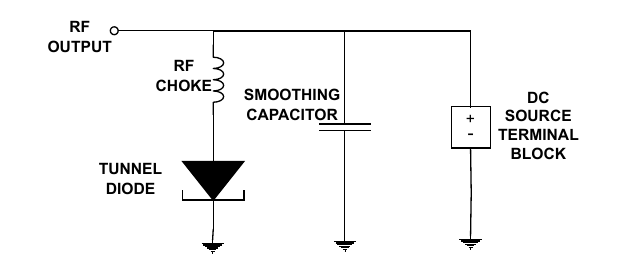}
    \caption{Board schematic for the tunnel diode reflection amplifiers/oscillators.}
    \label{fig:board_schematic}
\end{figure}

\begin{table}[htbp!]
    \centering
    \begin{tabular}{|c|c|}
    \hline
         Parameter & Value \\
    \hline\hline
        Tunnel Diode & \textit{AI101E} \\
        RF Choke & 18 nH\\
        Smoothing Capacitor & 0.1 $\mu$F \\
        Board Substrate & FR4 (4-layer) \\
    \hline
    \end{tabular}
    \caption{Tunnel Diode Board Parameters}
    \label{tab:refl_amp_board_params}
\end{table}

\begin{figure}
    \centering
    \includegraphics[width=0.85\linewidth]{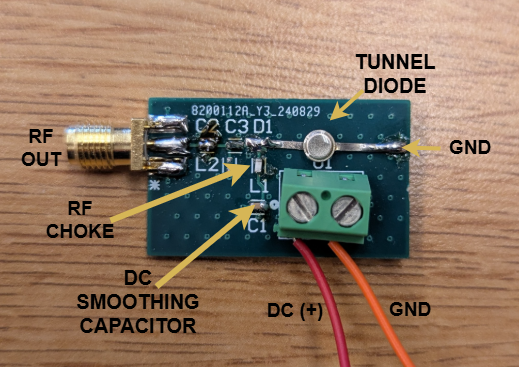}
    \caption{Fabricated tunnel diode board.}
    \label{fig:board_annotated}
\end{figure}

The schematic in Fig. \ref{fig:board_schematic} shows the layout of the tunnel diode boards, and Table \ref{tab:refl_amp_board_params} lists their parameters. The parameters of the AI101E tunnel diode used can be found at \cite{russian_tunnel_diode}. Each tunnel diode board, as pictured in Fig. \ref{fig:board_annotated}, was originally designed to be reflection amplifiers for \cite{Saetia_APSK}. Because of the circuit setup of the DC-biasing and tunnel diode portions of the board, these reflection amplifier boards also can operate as oscillators. The dual-functionality of tunnel diodes to serve as the basis of reflection amplifiers and oscillators had been noted by Eid et al. \cite{Eid_Tunnel_Diode} and Varshney et al. \cite{Varshney_TUnnelScatter}. It is the intention of this work for these tunnel diode boards to serve as both reflection amplifier backscatter modulators and harmonics-generators.

Interestingly, all boards were designed to be of the same specifications in Table \ref{tab:refl_amp_board_params}. Any differences in these boards came from the manufacturing and fabrication processes of their components and how they were even assembled on the board. These slight differences will cause different identifiable responses, as will be shown in the later sections.

\section{Cabled Measurements}

\subsection{Cabled Measurement Setup}
\begin{figure}
    \centering
    \includegraphics[width=0.95\linewidth]{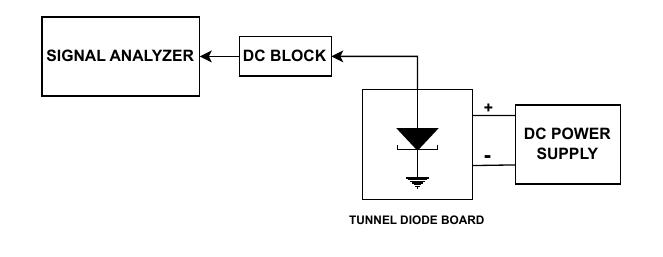}
    \caption{Cabled measurements setup.}
    \label{fig:TD_measurement_setup}
\end{figure}

Shown in Fig. \ref{fig:TD_measurement_setup}, an Agilent N9000A Signal analyzer, with a DC block, is connected to the tunnel diode board under test. A DC power supply biased the tunnel diode at various voltages. The measurements setup parameters for the signal analyzer and the DC power supply are listed in Table \ref{tab:measurements_setup}.

\begin{table}[htbp!]
    \centering
    \begin{tabular}{|c|c|}
    \hline
         Parameter & Value \\
    \hline\hline
        Number of Points & 10001\\
        Start Frequency & 50 kHz \\
        End Frequency & 3 GHz \\
        Resolution Bandwidth & 1 MHz \\
        Biasing Voltage Sweep & 3-300 mV (1 mV increments)\\
    \hline
    \end{tabular}
    \caption{Over-the-cable measurement setup parameters.}
    \label{tab:measurements_setup}
\end{table}

Notably, for all these measurements, no interrogation or injection-locking signal is sent to the boards. These boards were simply biased by the DC power supply, and it was predicted that the voltages needed to bias the diodes in the negative differential resistance (NDR) regions will lead to the appearance of harmonic content in their measured spectra. This behavior is evident as will be shown in the next section that showcases these boards' measurements at 200 mV, specifically.

\subsection{DC Voltage, Current, and Power Consumption}

    \begin{figure*}[htbp!]
        \centering
        \begin{subfigure}[b]{0.95\textwidth}
            \includegraphics[width=0.95\linewidth]{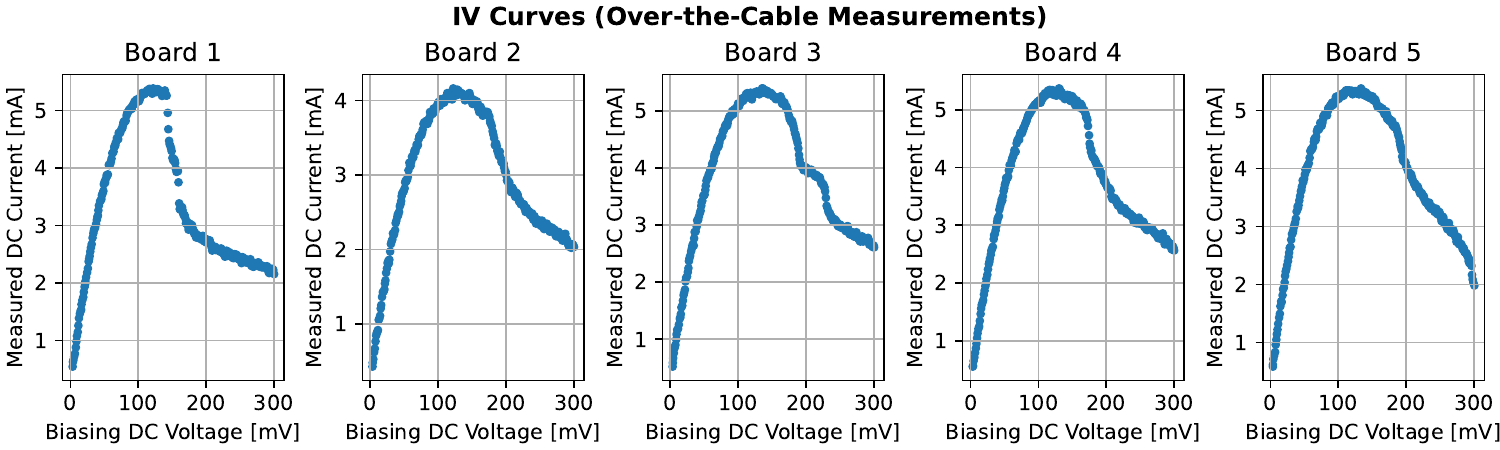}
            \caption{}
            \label{fig:IV_curves_cabled}    
        \end{subfigure}
        \hfill
        \begin{subfigure}[b]{0.95\textwidth}
            \includegraphics[width=0.95\linewidth]{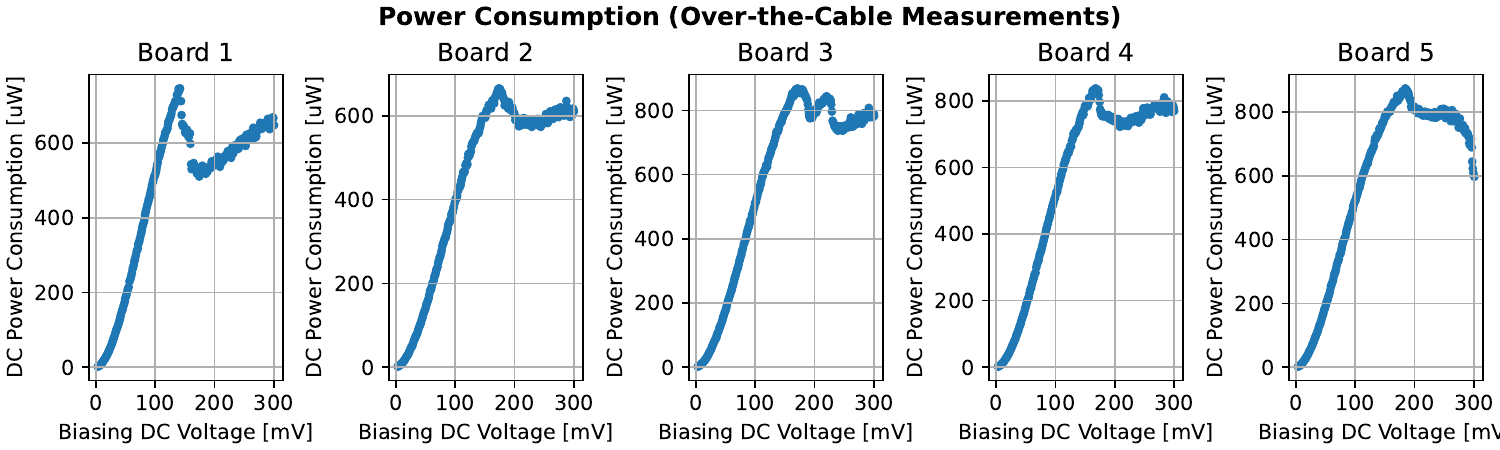}
            \caption{}
            \label{fig:power_consumption_cabled}    
        \end{subfigure}
        \hfill
        \caption{Top figure plots the measured DC current drawn by the tunnel diodes in relation to the measured applied DC biasing voltages. Bottom figure plots the measured DC power consumed by the tunnel diodes.}
    \end{figure*}

Fig. \ref{fig:IV_curves_cabled} shows the measured DC current as each board was swept through different biasing voltages. Likewise, Fig. \ref{fig:power_consumption_cabled} demonstrates the calculated DC power consumption for each board from multiplying the measured DC current and biasing voltage values. Device-dependent, the tunnel diodes started to enter their negative differential resistance regions at around 100 mV, then fully into their NDR regions when $\geq$150 mV  of biasing voltage was applied, evident in Fig. \ref{fig:IV_curves_cabled} by the sharp negative slope in the current-voltage (IV) curves. For  Fig. \ref{fig:power_consumption_cabled}, in the NDR regions, the curves plateaued after a steep rise prior to entering the NDR regions.

\subsection{Harmonic Behavior at 200 mV}

\begin{figure*}[htbp!]
    \centering
    \includegraphics[width=0.95\linewidth]{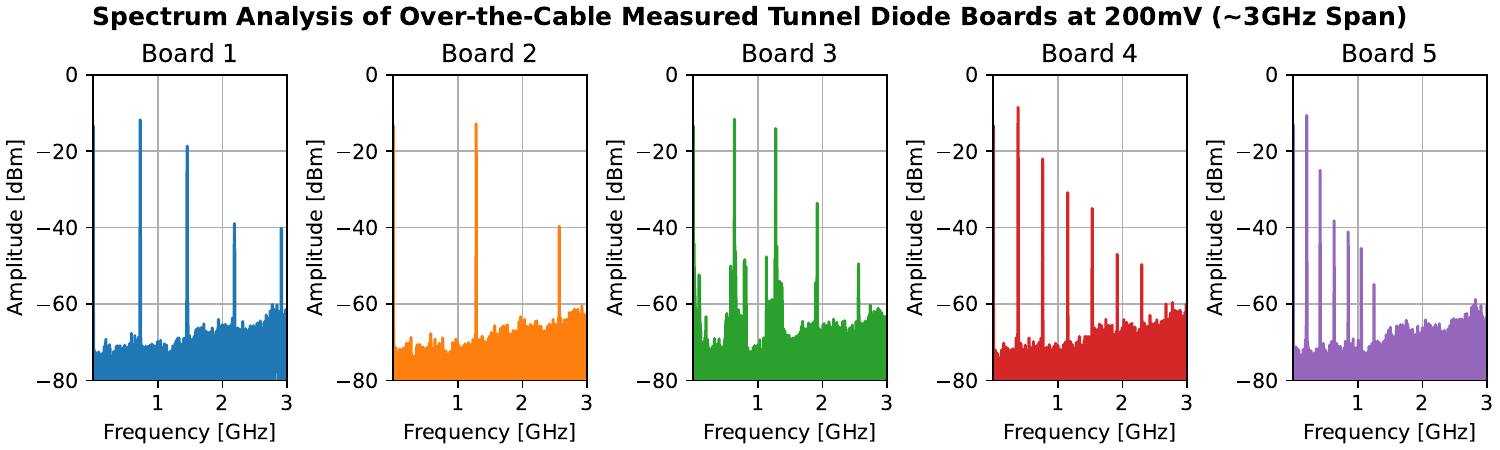}
    \caption{The measured spectrum of each tunnel diode board over about a 3 GHz span.}
    \label{fig:AI101E_Five_200mV}
\end{figure*}

At 200 mV, the captured spectrum of the five \textit{AI101E} reflection amplifier boards are plotted in Fig. \ref{fig:AI101E_Five_200mV}. Note, these output spectrum measurements, along with the ones in Fig. \ref{fig:AI101E_Color_Maps}, did not exclude the combined cable and DC block insertion losses, which was at most 0.2 dB and negligible. Shown by this figure and the fundamental frequencies listed in Table \ref{tab:fund_freqs}, each board had varying fundamental frequencies and different number of harmonics within the measured 3 GHz span. Note, the fundamental frequency is defined as the highest-amplitude, non-DC frequency content. Each board's fundamental frequency appeared to be at least above -15 dBm.

\begin{table}[htbp!]
    \centering
    \begin{tabular}{|c|c|c|}
    \hline
        Board & Fundamental Frequency (MHz) & Amplitude (dBm)  \\
    \hline
         1 & 727.2 & -11.80 \\
         2 & 1283 & -12.87  \\
         3 & 638.7 & -11.65 \\
         4 & 384.9 & -8.559 \\
         5 & 210.0 & -10.65 \\
    \hline
    \end{tabular}
    \caption{Fundamental Frequencies at 200 mV}
    \label{tab:fund_freqs}
\end{table}

\subsection{Spectrum-Voltage Signatures}
\begin{figure*}[htbp!]
    \centering
    \includegraphics[width=0.95\linewidth]{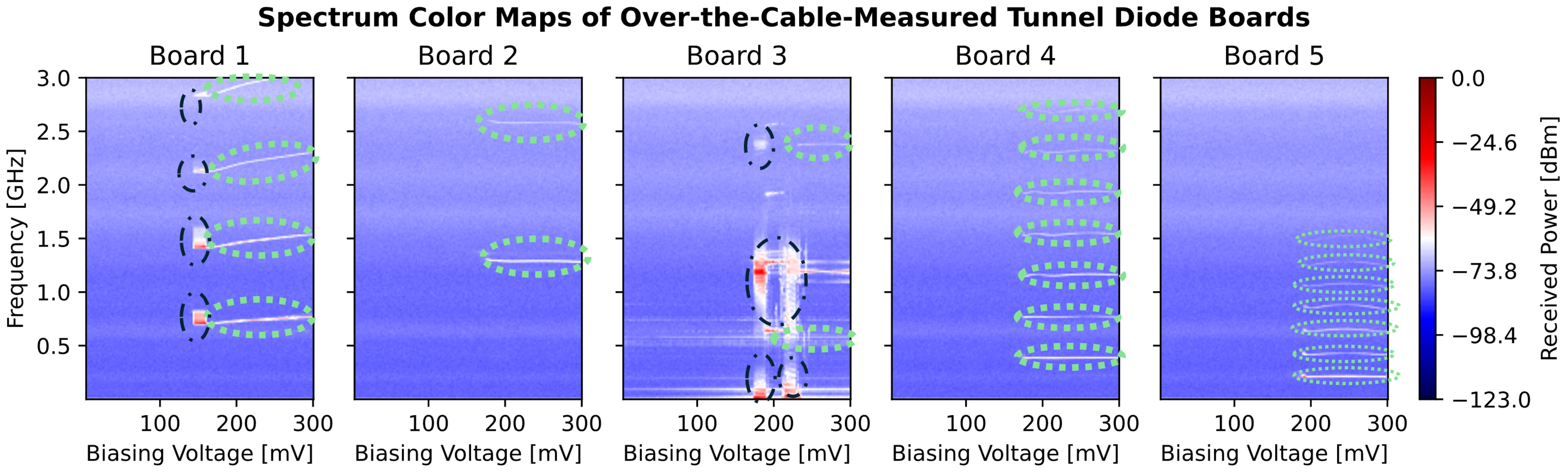}
    \caption{Performing a frequency and voltage sweep, a spectrum color map for each board is created. The pattern of warm-colored stripes show the harmonic signature for each board across a range of voltages. The green dotted circles highlight the stable harmonics, while the black dash-dot loops highlight the jittery harmonic response as the tunnel diodes enter their NDR regions.}
    \label{fig:AI101E_Color_Maps}
\end{figure*}

To investigate what happens when different voltages are applied to the tunnel diode boards, the measured spectrum from the signal analyzer for each biasing voltage sweep value are plotted as a color-map in Fig. \ref{fig:AI101E_Color_Maps}. As seen in that figure, there are lighter-colored "stripes" for where the tunnel diodes are resonating. Up to 150-180 mV, depending on the specific board, the diodes were not biased into their negative differential resistance regions; thus, the boards were not resonating. But then as the boards are biased closer to the biasing range of their NDR regions, the spectral response became jittery before stabilizing.

\begin{figure*}
    \centering
    \includegraphics[width=0.95\linewidth]{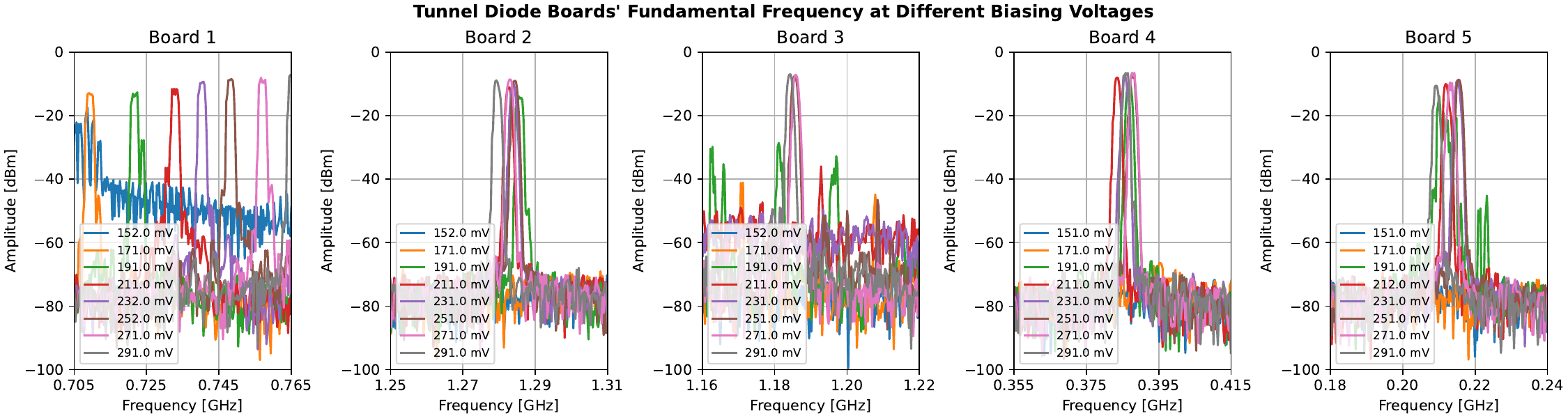}
    \caption{Shifting the fundamental frequency by applying different biasing voltages within the tunnel diodes' NDR regions.}
    \label{fig:spectrum_walk}
\end{figure*}

A close-up of each board's fundamental frequency in Fig. \ref{fig:spectrum_walk} demonstrates that the different biasing voltages can tune the fundamental frequency location by biasing them with different values within their diodes' negative differential resistance regions. Board 1 has the widest tunable range of about 60 MHz, while Board 3 has the narrowest tunable range of about 1 MHz. Varshney and Corneo \cite{Varshney_Tunnel_Emitter} noted similar frequency tuning for the \textit{AI101D} tunnel diodes based on input biasing voltage changes.

    \subsection{DC-to-RF Efficiency}
        \begin{figure*}[htbp!]
        \centering
        \begin{subfigure}[b]{0.95\textwidth}
            \includegraphics[width=0.95\linewidth]{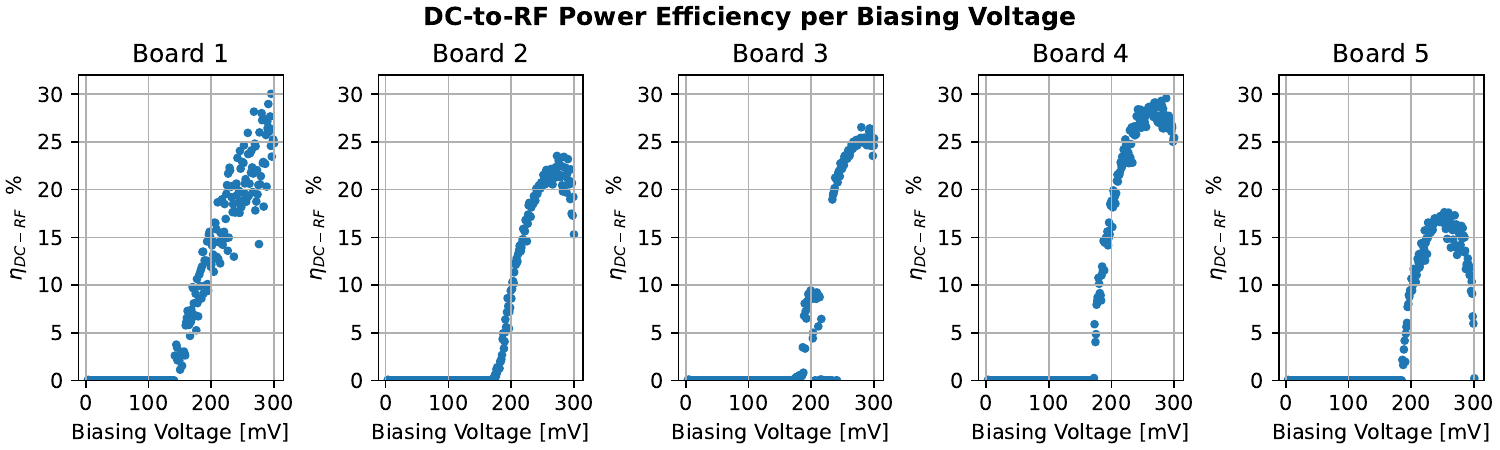}
            \caption{$\eta_{DC-RF}$ in relation to biasing voltage.}
            \label{fig:RF_DC_volt_x_axis}    
        \end{subfigure}
        \hfill
        \begin{subfigure}[b]{0.95\textwidth}
            \includegraphics[width=0.95\linewidth]{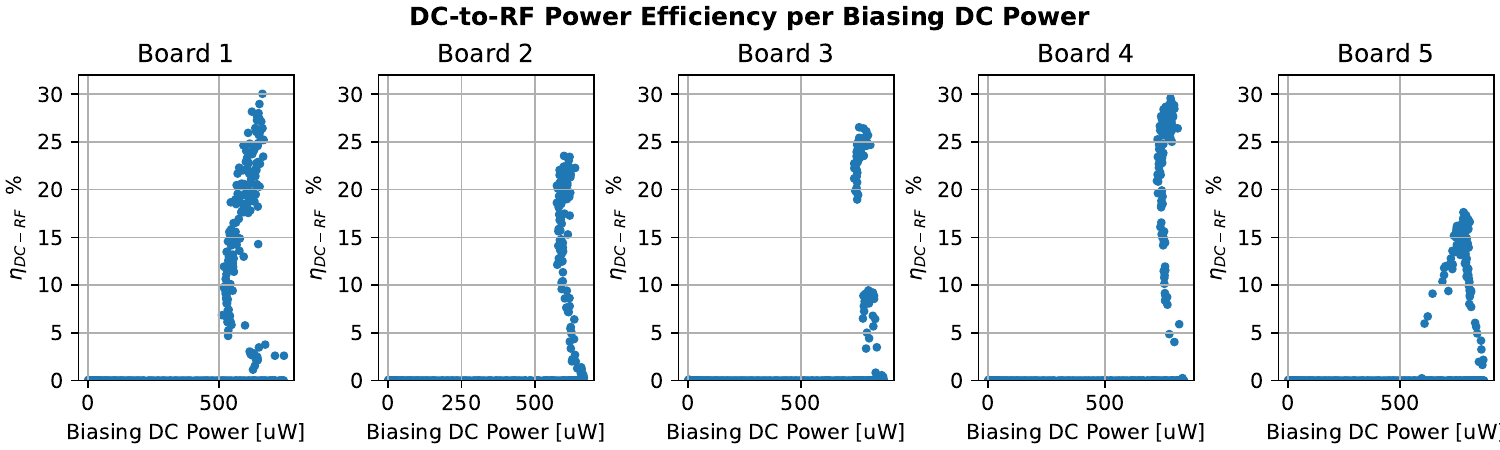}
            \caption{$\eta_{DC-RF}$ in relation to DC power consumed.}
            \label{fig:RF_DC_power_x_axis}    
        \end{subfigure}
        \hfill
        \caption{The DC-to-RF efficiency for each board is plotted and Board 1 demonstrated the greatest efficiency of the five boards at 30.04$\%$.}
    \end{figure*}

    A figure-of-merit for these tunnel diode oscillators are their DC-to-RF efficiency ($\eta_{DC-RF}$). This metric indicates how well these oscillators can convert DC power from the bias supply to an output RF signal. Since these oscillators are intentionally producing harmonics, $\eta_{DC-DC}$, defined in \eqref{RF_DC_efficiency}, simply is a ratio of the peak power output at a single frequency to the measured DC power consumed by the oscillator board. Fig. \ref{fig:RF_DC_volt_x_axis} and Fig. \ref{fig:RF_DC_power_x_axis} plot the DC-to-RF efficiency in relation to the applied biasing voltage ($V_{bias}$) and DC power consumed ($P_{DC}$). Demonstrated in Fig. \ref{fig:RF_DC_volt_x_axis}, the efficiency increased as the tunnel diodes were further biased into their NDR regions, peaked, and then decreased. Table \ref{tab:efficiency values} lists the maximum efficiency values for each board, with Board 1 having the best efficiency of 30$\%$ out of the five boards.
    
        \begin{equation}\label{RF_DC_efficiency}
            \eta _{DC-RF} = \frac{P_{peak,out}}{P_{DC}}
        \end{equation}

    \begin{table}[htbp!]
    \centering
    \begin{tabular}{|c|c|c|c|}
    \hline
        Board & Maximum $\eta_{DC-RF}$ ($\%$) & $P_{DC}$ ($\mu W$) & $V_{bias}$  (mV) \\
    \hline
         1 & 30.04 & 664.6 & 295.9\\
         2 & 23.53 & 597.7 & 273.1   \\
         3 & 26.53 & 758.5 & 279.9 \\
         4 & 29.58 & 779.6 & 287.4\\
         5 & 17.63 & 782.3 & 252.9 \\
    \hline
    \end{tabular}
    \caption{DC-RF Efficiency Values}
    \label{tab:efficiency values}
\end{table}

To the best of the authors' knowledge, these efficiency values are promising in relation to other tunnel diode oscillator designs. Eid et al \cite{Eid_Tunnel_Diode} achieved an output of a round -40 dBm with an input of 20 $\mu W$ ($\sim$43 dBm) for an efficiency of about 0.5$\%$ with their designed tunnel diode oscillator. At microwave frequencies, Wang \cite{Wang2012} designed a tunnel diode oscillator that outputted -10.17 dBm at 618 MHz (comparable to the RF output power levels of this work) and was highly efficient (53$\%$) due to the low 0.18 mW of power-consumption of his tunnel diode of choice. He did note that the power dissipation along his DC decoupling resistor caused that design's efficiency to drop to 2.8$\%$ \cite{Wang2012}. Spudat et al. \cite{spudat_RTD} listed the calculated efficiencies for their works and other comparable RTD oscillator designs at millimeter-wave (mmWave) and past mmWave frequencies to be at most $14.7 \%$, achieved by \cite{Cornescu_RTD}. These high frequency oscillator designs can output high output powers, such as greater than half a milli-watt, but consumes higher input power than this work, hence the low efficiencies. Also, some of these efficiencies were calculated with the resonant antenna structures on them, unlike this work's reported efficiencies that did not measure the power values with an antenna on the oscillator boards.

\section{Wireless Measurements Verification}
To verify the over-the-cable results and assess the possibility of detecting harmonic signatures from tunnel diodes, wireless measurements were completed in an anechoic chamber. 

    \subsection{Wireless Measurement Setup}
    
    \begin{figure}[htbp!]
        \centering
        \includegraphics[width=0.85\linewidth]{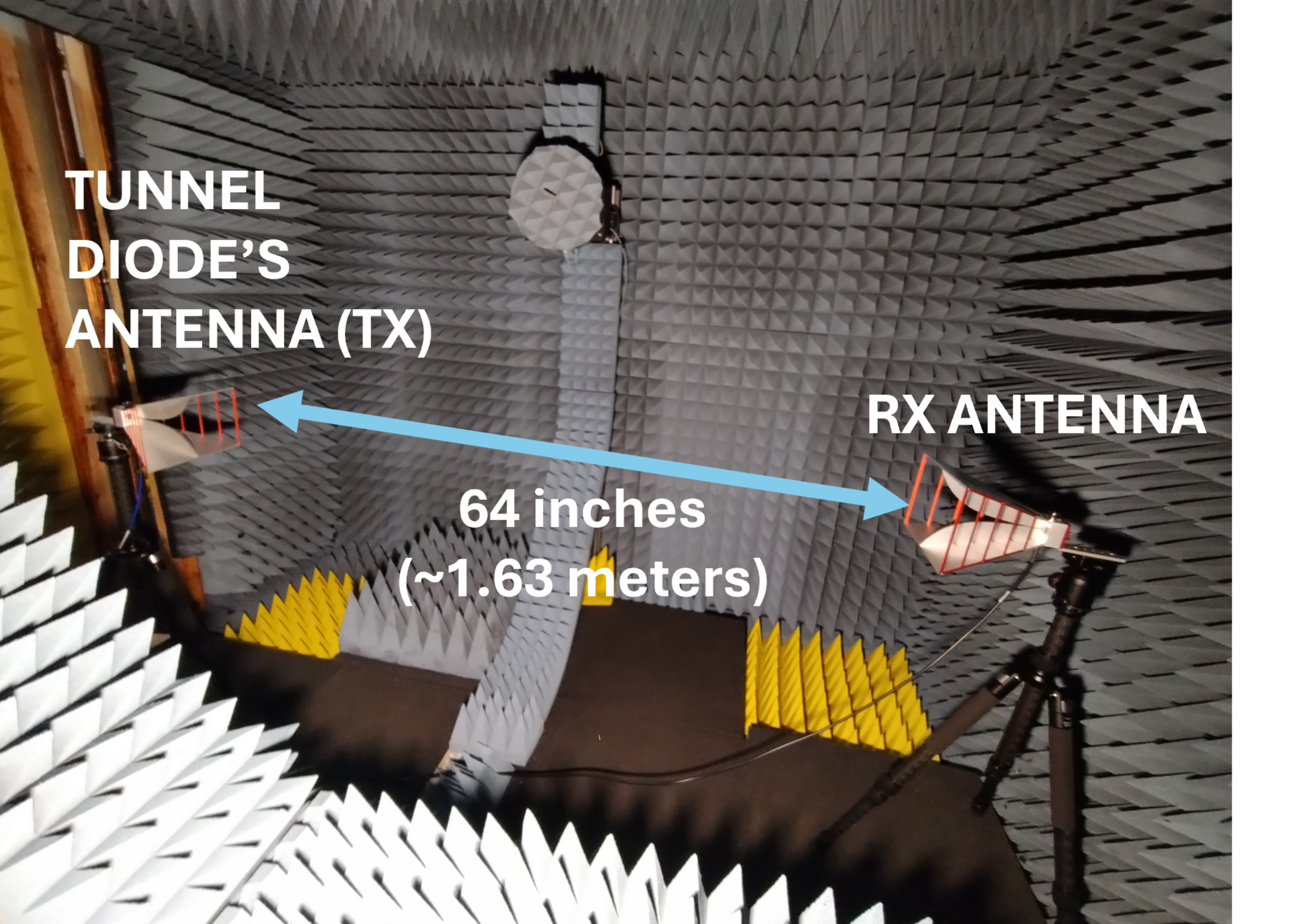}
        \caption{Antennas setup.}
        \label{fig:ant_setup}
    \end{figure}

    \begin{figure}
        \centering
        \includegraphics[width=0.85\linewidth]{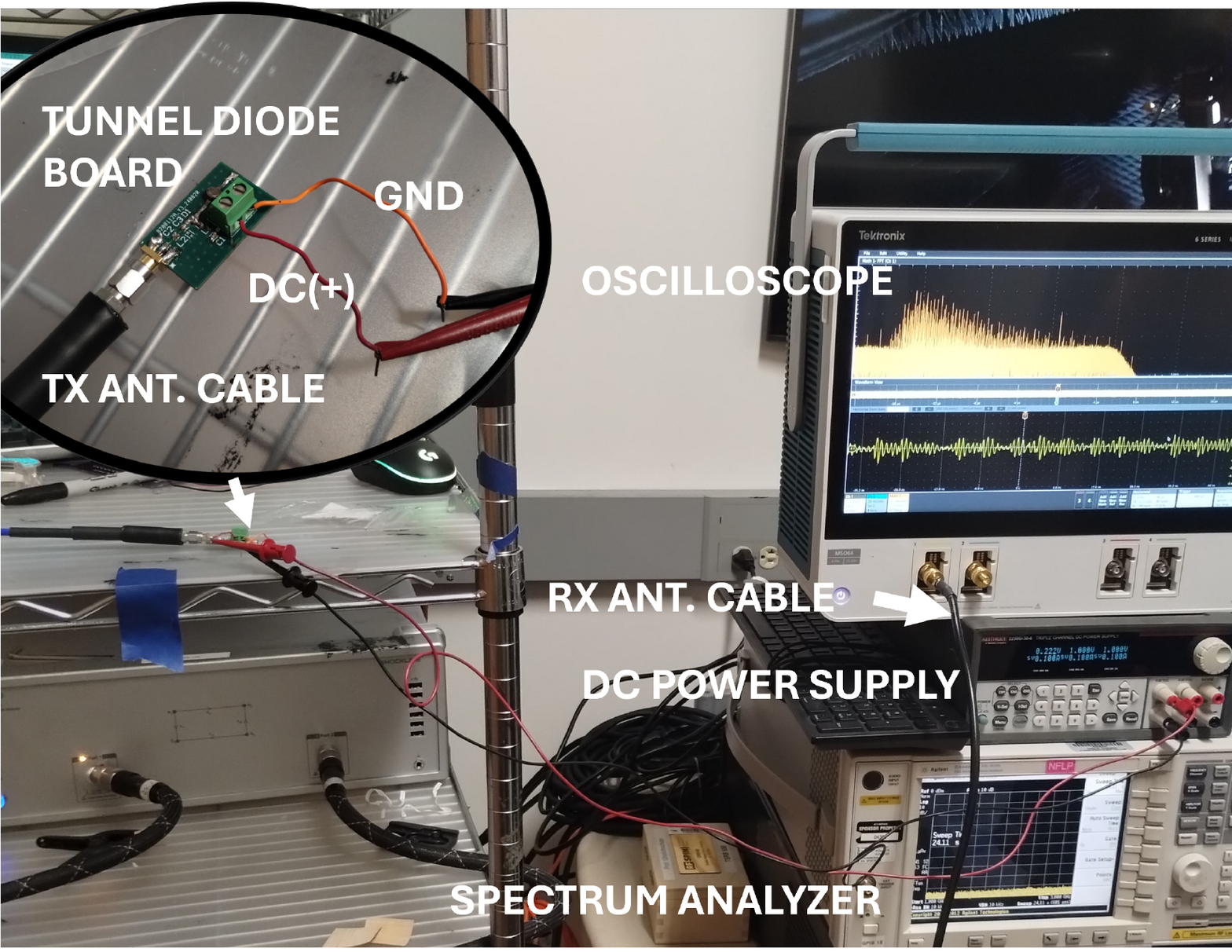}
        \caption{Measurement equipment setup.}
        \label{fig:wireless_equip_setup}
    \end{figure}

    Two wide-band and high-gain horn antennas were spaced 64 inches or 1.63 meters apart in an anechoic chamber, shown in Fig. \ref{fig:ant_setup}. These antennas are specifically rated to be 50 ohms, operate from 1-18 GHz, and provide over 7 dBi of gain past 1 GHz \cite{antenna_specs}. One antenna was connected to the tunnel diode board, the transmitter (TX). The other antenna was connected to either a spectrum analyzer or an oscilloscope, the receiver (RX). Due to the limited physical space of the chamber, 1.63 meters is the furthest the two antennas were spaced from each other. Shown in Fig. \ref{fig:wireless_equip_setup}, the tunnel diode boards, the spectrum analyzer, and oscilloscope were all placed outside the anechoic chamber during testing. The TX cable connecting the tunnel diode board to the TX antenna had at most 3.68 dB of insertion loss, and the RX cable connecting the RX antenna had at most 2.24 dB of insertion loss. Note, these losses were not accounted for in the following captured spectra and transient measurements.

    Two sets of measurements were taken in the frequency and time domains with, respectively, an Agilent E4446A spectrum analyzer and a Tektronix MSO64 oscilloscope that has 8 GHz bandwidth channels and a real-time sampling rate of 25 GS/s. For the wireless spectrum measurements, the following parameters in Table \ref{tab:wireless_measurements_setup} were used to set-up the spectrum analyzer and the DC power supply. For the time-domain/transient measurements taken with the oscilloscope, the horizontal span and trigger were adjusted manually to best capture the waveform.

    \begin{table}[htbp!]
        \centering
        \begin{tabular}{|c|c|}
        \hline
             Parameter & Value \\
        \hline\hline
            Number of Points & 8190\\
            Start Frequency & 50 kHz \\
            End Frequency & 3 GHz \\
            Biasing Voltage Sweep & 0-307 mV (1 mV increments)\\
        \hline
        \end{tabular}

        \caption{Wireless measurement setup parameters.}
        \label{tab:wireless_measurements_setup}
    \end{table}

    \subsection{Wireless-Measured Spectrum Results}
    
    \begin{figure*}[htbp!]
        \centering
        \includegraphics[width=0.95\linewidth]{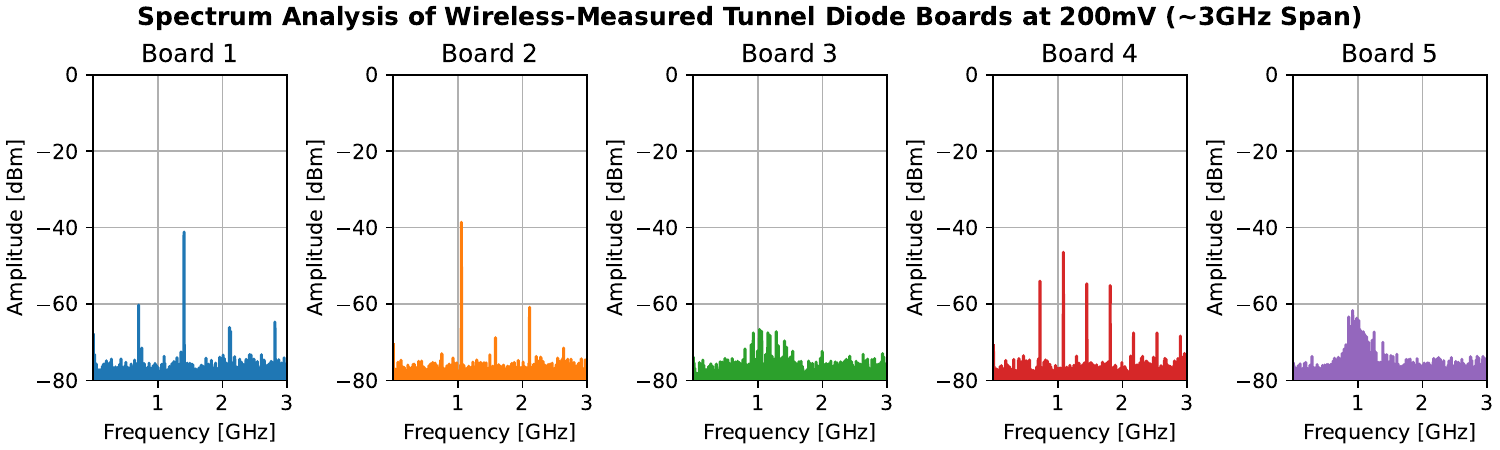}
        \caption{The measured spectrum of each tunnel diode board over about a 3 GHz span.}
        \label{fig:AI101E_Five_Wireless_200mV}
    \end{figure*}

    \begin{figure*}[htbp!]
        \centering
        \includegraphics[width=0.95\linewidth]{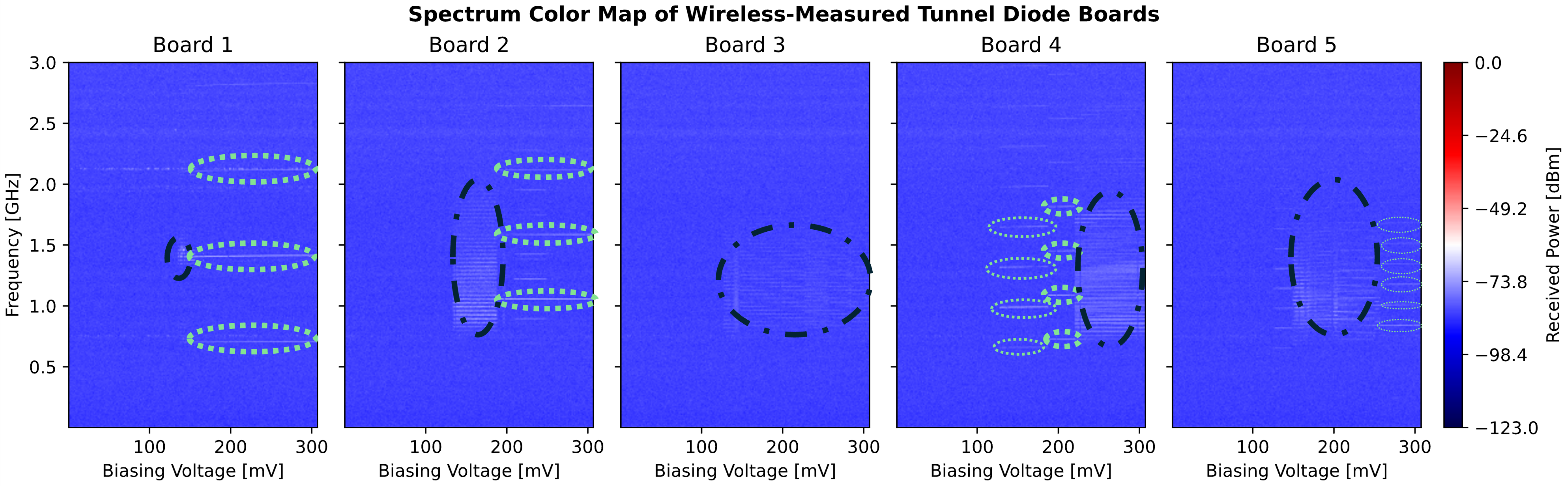}
        \caption{The measured spectrum of each tunnel diode board over about a 3 GHz span. The green dotted loops emphasize the stable harmonics. While the black dash-dot loops highlight the wide-band, jittery/unstable harmonic resonance regions.}
        \label{fig:Color_Map_Wireless}
    \end{figure*}

    \begin{figure*}[htbp!]
        \centering
        \includegraphics[width=0.95\linewidth]{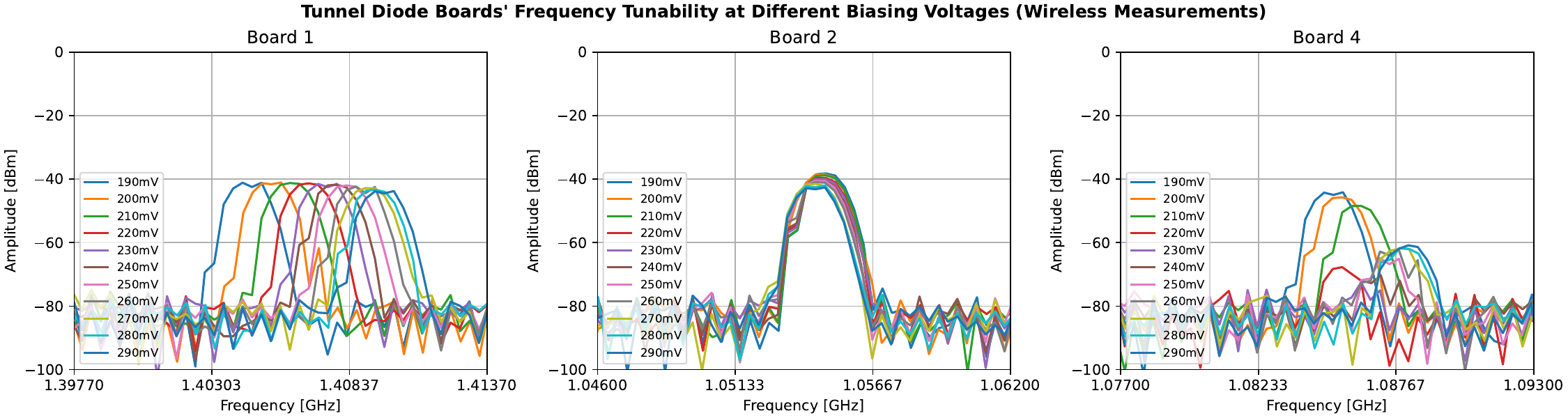}
        \caption{The tunability of the harmonics of the tunnel diode boards when measured wireless is dependent on the biasing voltage as demonstrated by the results of the boards' wireless measurements.}
        \label{fig:Spectrum_Walk_Wireless}
        
    \end{figure*}
    
    Fig. \ref{fig:AI101E_Five_Wireless_200mV} shows the wireless measured harmonic signatures of the tunnel diode boards at 200 mV. Boards 1, 2, and 4 from the wireless measurements have notable harmonic signatures at 200 mV. Boards 3 and 5 do demonstrate frequency-producing behavior at 200 mV, but not distinct harmonics as their spectrum content is spread/clustered over several frequencies. The differences in the measured over-the-cable and wireless results can be possibly due to impedance mismatch between those specific boards and the antenna and cable and/or the antenna's impedance modifying the overall resonant behavior of the boards when the antenna was loaded onto the boards. Fig. \ref{fig:Color_Map_Wireless} demonstrates the harmonic signature behavior of the boards over several biasing voltages. Because of path loss and cable losses, the receive strength of the harmonic signatures was not as strong as the measured over-the-cable signatures captured in Fig. \ref{fig:AI101E_Color_Maps}. To aid with visualization, the green dotted loops in Fig. \ref{fig:Color_Map_Wireless} highlights the stable harmonics of the boards, while the black dash-dot loops mark the jittery-wide-band spectrum  behavior, demonstrated for example by Boards 3 and 5 in Fig. \ref{fig:AI101E_Five_Wireless_200mV}. As expected, Boards 1 and 2 have stable, distinct harmonic signatures when biased in their negative differential resistance regions. The jittery, wide-band behavior demonstrated by Boards 3 and 5, even in their NDR regions, could be because of impedance mismatch with the antenna and cable as the biasing voltage on the board changes the board's impedance.
    
    It should be noted that because the antennas have below 0 dBi of gain at frequencies below about 700 MHz \cite{antenna_specs}, sub-1 GHz fundamental components will not appear as strong as they did in the over-the-cable measurement results. Table \ref{tab:fund_freqs_wireless} shows the predicted fundamental frequency locations based on these wireless measurements and the location of the visible second harmonics. Also, the harmonics locations for the boards measured wireless occurred at different locations than those that were measured over-the-cable due to the change in testing setup with attaching the antenna and longer cable with different impedances. Such as shown in Fig. \ref{fig:spectrum_walk} with the over-the-cable measurements, Fig. \ref{fig:Spectrum_Walk_Wireless} demonstrates the tunability of the tunnel diode boards with the wireless measurement results. Visually inspecting Fig. \ref{fig:Spectrum_Walk_Wireless}, Board 1 demonstrated the best tunable range at about 5 MHz. It should be noted the center frequencies in that figure are of the strongest frequency component in the captured spectra of Fig. \ref{fig:AI101E_Five_Wireless_200mV}, so they are not always the fundamental frequencies as were used in Fig. \ref{fig:spectrum_walk}.

    \begin{table}[htbp!]
        \centering
        \begin{tabular}{|c|c|}
        \hline
            Board & Predicted Fundamental Frequency (MHz)\\
        \hline
             1 & 702.9  \\
             2 & 527.4  \\
             3 & N/A   \\
             4 & 361.9  \\
             5 & N/A  \\
        \hline
        \end{tabular}
        \caption{Predicted Fundamental Frequencies (Wireless) at 200 mV from the Wireless Measurements}
        \label{tab:fund_freqs_wireless}
    \end{table}

    \subsection{Time Domain Measurements}
    \begin{figure*}[htbp!]
        \centering
        \includegraphics[width=0.95\linewidth]{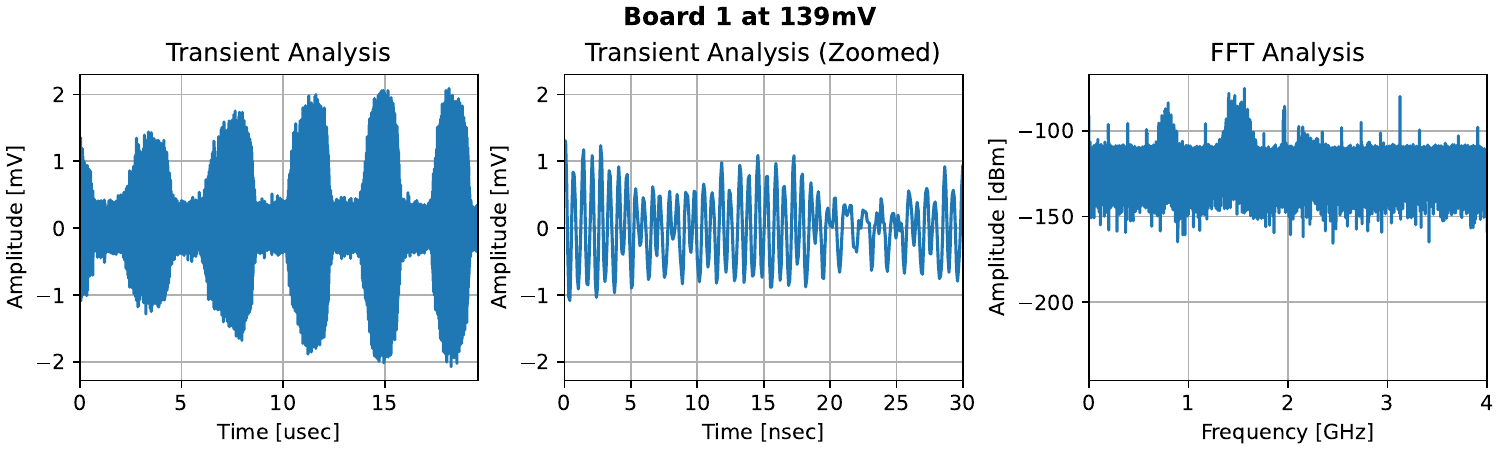}
        \caption{The captured transient data of Board 1 biased at 139 mV, and the oscilloscope's FFT analysis on the captured data.}
        \label{fig:Board_1_Time_Domain_139mV}
    \end{figure*}

    \begin{figure*}[htbp!]
        \centering
        \includegraphics[width=0.95\linewidth]{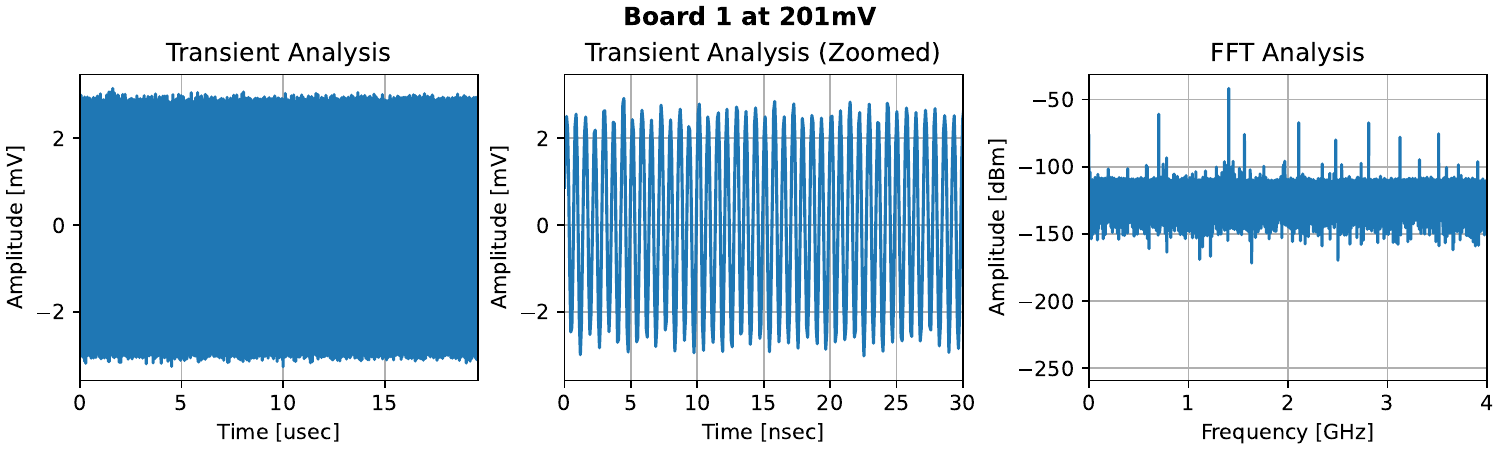}
        \caption{The captured transient data of Board 1 biased at 201 mV, and the oscilloscope's FFT analysis on the captured data.}
        \label{fig:Board_1_Time_Domain_201mV}
    \end{figure*}

    \begin{figure*}[htbp!]
        \centering
        \includegraphics[width=0.95\linewidth]{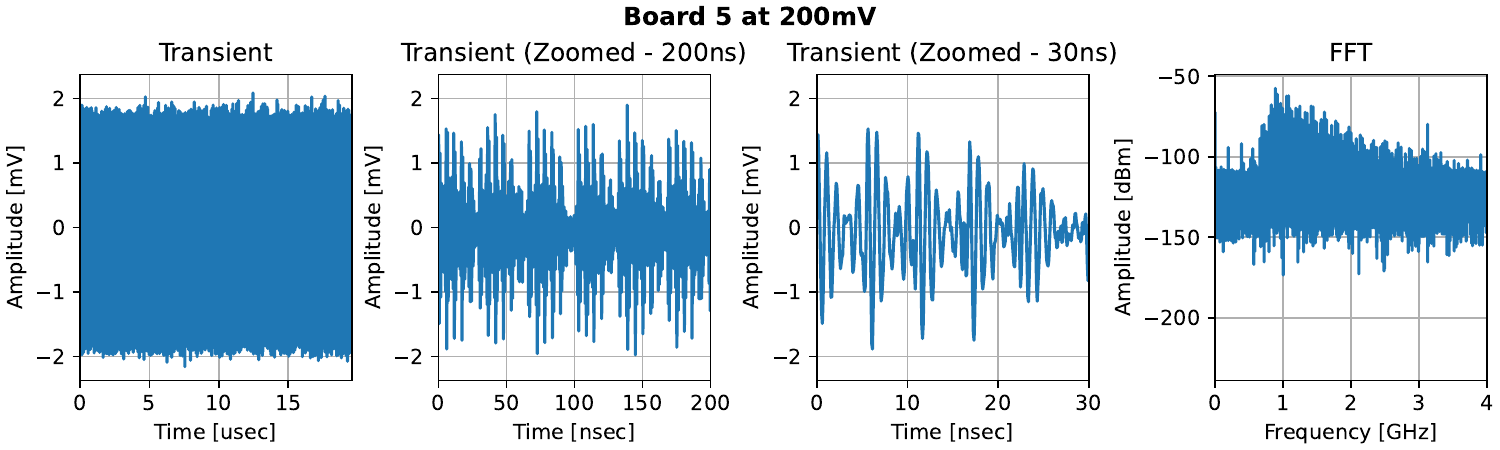}
        \caption{The captured transient data of Board 5 biased at 200 mV, and the oscilloscope's FFT analysis on the captured data..}
        \label{fig:Board_5_Time_Domain_200mV}
    \end{figure*}
    
    Using an oscilloscope, the wireless outputs from the tunnel diode boards were captured and measured when the boards were producing relatively stable harmonics and when they were not. This work demonstrates the transient behavior of Board 1 at 139 mV and 201 mV in Fig. \ref{fig:Board_1_Time_Domain_139mV} and Fig. \ref{fig:Board_1_Time_Domain_201mV}, respectively. Shown in Fig. \ref{fig:Board_1_Time_Domain_201mV}, at 201 mV, Board 1 produced a relatively distinct sinusoid, with the Fast-Fourier-Transform (FFT) analysis from the oscilloscope, showing also distinct harmonics. This result is consistent with Board 1's behavior at 200 mV in Fig. \ref{fig:AI101E_Five_200mV}. At 139 mV, Board 1 was entering the negative differential resistance region, and the output of the board was "bursty," as shown in the left-most plot of Fig. \ref{fig:Board_1_Time_Domain_139mV}. Within each burst is a Gaussian pulse-like shape, hence the spread-out spectrum FFT results in the right most plot of Fig. \ref{fig:Board_1_Time_Domain_139mV}. Similarly, Board 5 at 200 mV, in Fig. \ref{fig:Board_5_Time_Domain_200mV}, demonstrated a Gaussian pulse-like output and also a spread-out FFT result. Again, this behavior is consistent with what was demonstrated by Board 5 in Fig. \ref{fig:AI101E_Five_200mV}. Interestingly, these bursts that Board 1 at 139 mV and Board 5 at 200 mV demonstrated suggest that the tunnel diode oscillator boards act as self-blocking or even super-regenerative oscillators \cite{Armstrong_Regenerative_Circuit} as the measured output ramps up and then decays to build up oscillation for a new burst. 

\section{Link Budgets and Reader and Harmonic Tag Hardware Design Considerations}
        A wide-band, harmonic RFID reader and tag with a harmonic tunnel diode oscillator have to be designed with special considerations in regards to power consumption of the tag, read-range, and frequency planning for wide-band operation. Preliminary link-budgets of the forward and reverse links, with tunnel diode Board 1 serving as the tag, are presented to help theorize possible ranges and provide insight on design choices for the reader and tag. The reverse and forward links are significant since they are related to read-range and power-up range, respectively.
    
    \subsection{Reverse Link}
    \begin{figure}[htbp!]
        \centering
        \includegraphics[width=0.95\linewidth]{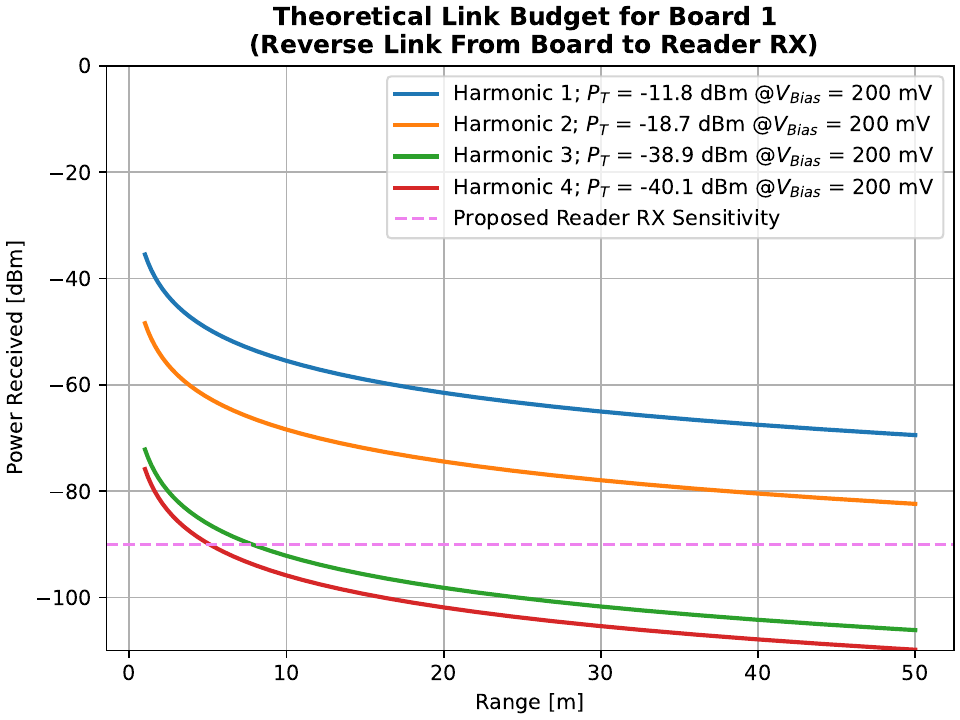}
        \caption{Theoretical reverse link budget for tunnel diode Board 1.}
        \label{fig:link_budget}
    \end{figure}
    
    The reverse link range is the range the information from a harmonic tag can be read by the reader's receiver. Using the over-the-cable measurement results and the power transmitted ($P_T$) by each harmonic for Board 1 at 200 mV, a theoretical one-way link budget between the tunnel diode board, acting as a tag, and a receiving RFID reader (RX) is calculated and shown in Fig. \ref{fig:link_budget}. The purpose of this link budget is to estimate the range of harmonic detection and determine specific design parameters for future reader receiver design to improve sensing/detection range. Note, the transmitted signal strength of the reverse link (from tag to reader) is not directly affected by the forward link (from tag to reader) since no backscattering is involved.
    
    In this link budget, the antennas for both the board and reader are assumed to be 3 dBi flat over the wide bandwidth that includes the fundamental and harmonic frequencies of Board 1. The reader sensitivity is set and proposed to match the rated sensitivity or lowest detection level of the Impinj R700 reader at -90 dBm \cite{Impinj_RFID_Reader}. As shown by doing this link budget, the fundamental (first) and second harmonics can be detected by a reader of that sensitivity past 50 meters. The third and fourth harmonic would not be detected by the reader for more than about 5.5 meters. Better reader sensitivity is needed to detect these later harmonics past 5 meters. Another important factor, indicative by this link budget, is to have a wide-band antenna. One of higher gain can help improve the power received at the reader. The reader must be designed to be wide-band and have low minimum-power detection sensitivity level to detect the higher order harmonics.

    \subsection{Forward Link}
    \begin{figure}[htbp!]
        \centering
        \includegraphics[width=0.95\linewidth]{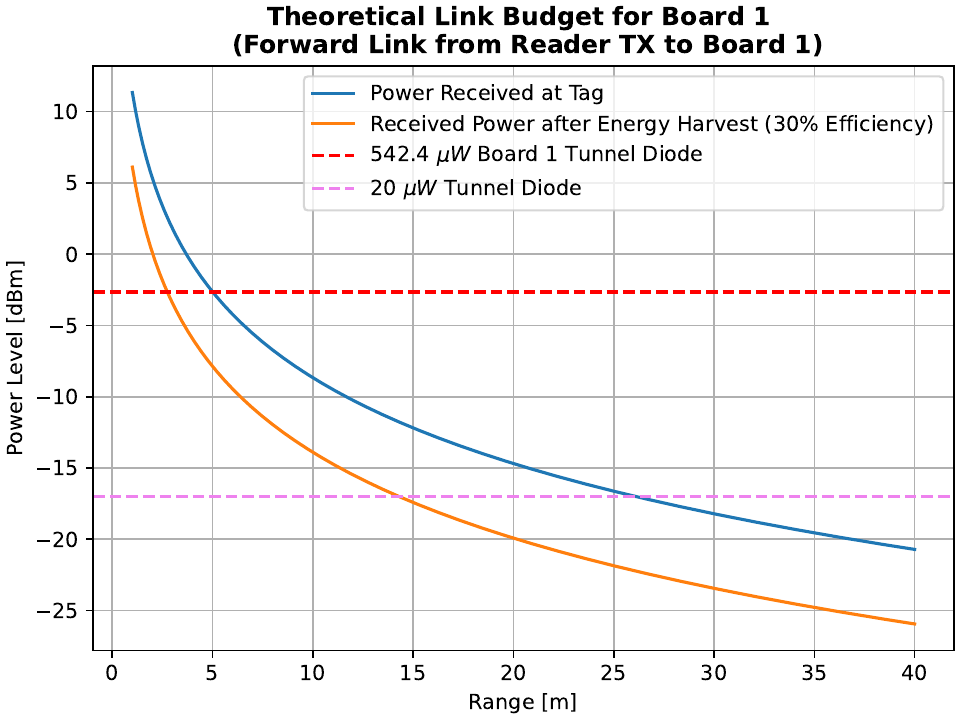}
        \caption{Theoretical forward link budget for tunnel diode board 1.}
        \label{fig:link_budget_forward}
    \end{figure}
    
    The forward link, from reader's transmitter (TX) to tag, is significant to determine the power-up range. Like the previous link-budget, the antennas are assumed to be wide-band and have flat 3 dBi gain. The power transmitted is a maximum 1 W, in compliance with the United States Federal Communications Commission (FCC). Fig. \ref{fig:link_budget_forward} shows the resulting forward link budget. The tag's rectification efficiency is taken to account with the orange curve in Fig. \ref{fig:link_budget_forward}. Even though rectification efficiency varies by input power \cite{Valenta_EH}, this link budget uses a constant 30$\%$ rectification efficiency when the tag is between 0 to 40 meters away from the reader transmitter.  

    As demonstrated by these results, a fully-passive design would be forward-range limited.  At 200 mV, Board 1's tunnel diode consumed 524.6 ($\mu W$) of power, which is significantly higher than the tunnel diode power consumption shown in \cite{Eid_Tunnel_Diode}, which consumed 20 $\mu W$. Thus, using the higher power-consuming tunnel diode would only allow the tag to be powered about 2.74 meters compared to 14.27 meters when using a lower-powered tunnel diode, such at the 20 $\mu W$ tunnel diode used in \cite{Eid_Tunnel_Diode}. The forward-range limitations can be resolved by using a semi-passive design that only biases the diode when the tag's ID is requested.

    \subsection{System Hardware Design Considerations}
    The main issues for this theorized harmonic RFID system are: diode's power-consumption and reader's sensitivity. As shown by the link budgets for the forward and reverse links, the system is more constrained with the forward-range compared to reverse-range due to the power-consumption of the tunnel diode. The tag would have better forward-range if it is semi-passive. Then, the reader will transmit a "wake-up" signal, and the tag can utilize a simple, passive wake-up receiver to then detect the wake-up signal and only bias the tunnel diode to emit a harmonic identifying signal without powering on the other components and processes of the tag. 
    
    The reader can be designed as traditional up-conversion transmit and down-conversion receive chains, with looser power and size constraints than the tag. Hence, the reader can use traditional mixers, amplifiers, and other active components. For improved receiver sensitivity, careful planning of the RF chain needs to be done to avoid desensitization by unwanted signals and only amplification of the wanted harmonic content, which can be challenging for wide-band ($>$500 MHz) operation. For the reverse-range, since the harmonic signature is wide-band, then the reader's receiver hardware will have to operate wide-band and need careful frequency planning to mitigate or filter out blockers or interferers that can saturate and desensitize the hardware. The wide-band operation also provides challenges in sourcing or designing the traditional RF components of a receive chain (such as amplifier, filters, etc.) to have strong performance and a flat response over a wide-band. Additionally, the antenna design for both the tag and reader's receiver needs to be designed to be wide-band to be able to transmit and capture the widely-spaced out harmonics from the tag.

\section{Harmonic Signatures and Applications}
Tunnel diodes can be included in future tag designs as part of their backscatter modulator circuitry. To actuate the generation of harmonic signatures from these diodes, semi-passive tags can periodically bias these tunnel diodes to wake-up and emit a harmonic signature, similar to Bluetooth beacons' advertisements.  These tags' tunnel diodes would only need to be biased in their negative differential resistance regions to produce distinct signatures. Shown by the measurements presented in this work, these signatures can be of high quality factor as those exhibited by the \textit{AI101E} diodes. There are two possible useful applications for having RFID reader receiving harmonic signatures from tunnel diodes: detecting RFID tags' presence in an area and identifying tags.

\subsection{Tag Detection}
The simplest use of harmonics is to aid RFID readers in determining if there are tags present in an area and even attempt to localize them. Using harmonics for detection have originated before the year 2000 with the use of "harmonic radar" and non-linear tags to track beetles \cite{beetle_tracking_harmonic_radar} and bees \cite{bee_tracking_harmonic_radar}, since insect-tracking requires mitigating the effects of strong clutter interference. Similarly, a RFID reader operates in an environment filled with clutter, other readers and transmitters, and strong self-interference. A reader can discover tags with their emitted harmonics and then go through with the standard communication protocols with the tags. The benefit of using harmonics for sensing tags, versus just listening for the tag's backscatter in the industrial-scientific-medicine (ISM) band, is that it mitigates self-interference issues as the harmonics are located away from the carrier-frequency in that band and loosens the need for the reader's receive chain to have sharp filter cut-offs when filtering out self-interference. Additionally, the use of harmonics that lie away from the carrier frequency band allows for improved localization of tags as demonstrated by Ma and Kan \cite{Indoor_Ranging_Harmonic_Tags} and Hui et. al \cite{CDMA_Harmonic_Localization}.

\subsection{Tag Identification}
Most chipped RFID tags have a serialized ID stored in memory. Mondal and Chahal \cite{ID_Encoding_Harmonic} developed a harmonic tag to encode the stored ID on a second harmonic of the interrogating or fundamental frequency. This work proposes, one-step further, the idea of using the harmonic signatures of different tags with tunnel diodes themselves to create \textit{memory-less, hardware-intrinsic signatures}. The previous sections of this paper presented the ability of tunnel diodes to generate "harmonic signatures." The presented signatures for each board are different from each other; though, uniqueness of signatures of course can not be guaranteed. Not shown in this work, because the board was damaged midway through testing, there was a board that had a really similar signature to Board 4 with similar harmonic locations, but at different measured strengths. Finer measurement resolution can distinguish the difference in the harmonics locations.

For a harmonic RFID reader, its receiver can perform several techniques to read the harmonic identifications from tags with tunnel diode oscillators. A software-defined receiver would be able to perform a FFT and thresholding. Or like a frequency-shift-key (FSK) detection scheme, the reader can perform correlation or matched filtering and envelope-detection \cite{Rappaport_Wireless_Comms} to detect known signatures. The process of learning and later identifying and verifying these harmonic signatures lead into the concept of RFID fingerprinting.

"Fingerprinting" RFID devices has been demonstrated through characterizing the minimum power at various frequencies needed to make tags respond \cite{Fingerprinting_Tags_Thompson} and also the backscatter responses (reflection coefficients) of the tags \cite{RCID_Fingerprinting}. Even tunnel diodes and their distinct IV curves have been proposed as the basis for physical-unclonable-functions (PUFs) \cite{Tunnel_Diode_PUFs}. Advanced classification methods for fingerprinting and identifying specific harmonic RFID tags will be investigated in future work. The color maps in Fig. \ref{fig:AI101E_Color_Maps} and Fig. \ref{fig:Color_Map_Wireless} can serve as rich input features for machine learning approaches. These representations can be used to create a convolutional neural network (CNN) system that not only classifies known devices but also performs open-set recognition to detect unknown devices, similar to the approach proposed in \cite{cnn-based-identification-lte}. Such an approach could extract discriminative features from these color maps while employing specialized loss functions like max-min loss to maximize inter-class separation while minimizing intra-class distances. This would enhance both the classification accuracy of known tags and detection capability for unknown tags, creating a more robust identification system.

Another exciting avenue to explore with these tunnel diodes oscillators' harmonic signatures and fingerprinting them is simultaneous identification and communications of multiple tags. For example, multiple tags in an area can be distinguished by their distinct and different signatures, and the reader can then choose to communicate with specific tags. For simultaneous communications, simple on-off-keying (OOK) by toggling the presence of these harmonic signatures or frequency-shift-keying (FSK) to adjust the harmonics locations can all be used to allow tags to simultaneously communicate, on different frequency bands that these signatures are located on, assuming each tag has a unique harmonic signature.

Another possible application for these tunnel diode-generated harmonic signatures is for security. Microwave tunnel diode circuits' are sensitive to "initial conditions" changes. As shown by this work, slight deviations in circuit component values, such as those that originate from manufacturing/fabrication processes, can cause seemingly identical circuit systems to produce distinct, different outputs. That same idea can also be applied to any malicious tampering done to tags with these tunnel diodes. It should be expected, then, for the harmonic signature to change if someone tries to place the tag on a different object or physically tamper with the tag.

\section{Conclusion}
This work presented the characterization of tunnel diodes, behaving as a harmonics-producing oscillator and demonstrated these diodes' capability to produce hardware-intrinsic "memory-less" and "backscatter-less" harmonic signatures for identification. Future studies, based on this current work, should investigate optimal tag antenna design to allow for wide-band sensing of these diodes' harmonic signatures wireless. A big limiting factor for using tunnel diodes for harmonic RFID would be the tag's antenna. Traditionally, a narrow-band antenna acts as a "filter" that dampens out-of-band harmonics from being emitted. As shown by the spacing of the harmonics in Fig. \ref{fig:AI101E_Five_200mV}, a wide-band antenna is needed by both the tag and reader to transmit and capture the almost random placement of these harmonics, respectively. If not a wide-band antenna, then an antenna that periodically resonates at a specific tunnel diode's harmonics can also be used. Future wide-band reader designs, with high sensitivity for detecting weaker higher-order harmonics, also should be investigated for detecting and processing wide-band harmonic signatures. Similar to other wide-band receivers, a harmonic RFID reader will have to deal with multiple non-RFID tag interfering transmitters that can desensitize the reader and need wide-band components, which bring challenges to the reader's design and frequency planning. Additional future wireless experimentation of tunnel diode tags should investigate and measure realistic sensing ranges for these tags and possible protocols for sensing and communicating with a group of these tags. For practicalness, methods to ensure harmonic tunnel diode RFID tags meet the regulations on ultra-wide-band wireless devices should also be investigated.

\section{Acknowledgments}
The authors of this paper want to thank Walter (Drew) Disharoon for aiding with the wireless measurements and Dr. Nima Ghalichechian and his mmWave Antennas and Arrays Laboratory for providing the anechoic chamber, antennas, and other supporting equipment for the wireless measurements. Thank you also to Zachary Brumbach and Dr. John Cressler for providing the spectrum analyzer for the wireless measurements.

\bibliographystyle{ieeetr} 
\bibliography{refs} 

\end{document}